\documentclass[prl,nobalancelastpage,nofootinbib,superscriptaddress,showpacs]{revtex4-2}

\usepackage{color,amsthm,amsmath,amsfonts,graphicx,bm}

\usepackage[utf8]{inputenc}
\usepackage{fullpage}
\usepackage[breakable]{tcolorbox}
\usepackage{xcolor}
\usepackage{algorithm}
\usepackage[noend]{algpseudocode}
\usepackage{wrapfig}

\usepackage{appendix}
\usepackage{amsmath}
\usepackage{amsthm}
\usepackage{amsfonts}
\usepackage{graphicx}
\usepackage{amssymb}
\usepackage{dsfont}
\usepackage{mathtools}
\usepackage{enumerate}
\usepackage{braket}
\usepackage{tikz}
\usetikzlibrary{quantikz}
\usepackage{wrapfig}
\usepackage{tikz-cd}

\newtheorem{definition}{Definition}
\newtheorem{theorem}{Theorem}
\newtheorem{corollary}{Corollary}

\newtheorem{example}{Example}

\newcommand{\dmat}[1]{{| #1 \rangle \langle #1 |}}

\usepackage{cleveref}

\DeclareMathOperator{\Tr}{\textnormal{Tr}}

\begin{document}

\preprint{APS/123-QED}
\title{DQC1-hardness of estimating correlation functions}
\author{Subhayan Roy Moulik}
\email{roymoulik@damtp.cam.ac.uk}
\affiliation{Department of Applied Mathematics and Theoretical Physics, University of Cambridge, UK}
\author{Sergii Strelchuk}
\email{sergii.strelchuk@cs.ox.ac.uk}
\affiliation{Department of Applied Mathematics and Theoretical Physics, University of Cambridge, UK}
\affiliation{Department of Computer Science, University of Oxford, UK}

\begin{abstract} 
Out-of-Time-Order Correlation function measures transport properties of dynamical \hbox{systems}. They are ubiquitously used to measure quantum mechanical quantities, such as \hbox{scrambling} times, criticality in phase transitions, and detect onset of thermalisation.
We characterise the \mbox{computational} complexity of estimating OTOCs over all eigenstates and show it is Complete for the One Clean Qubit model (DQC1). We then generalise our setup to establish DQC1-Completeness of \mbox{$N$-time} Correlation functions over all eigenstates. Building on previous results, the DQC1-Completeness of OTOCs and $N$-Time Correlation functions then allows us to highlight a dichotomy between query complexity and circuit complexity of estimating correlation functions. 

\end{abstract}

\maketitle

Correlation functions describe the statistical and dynamical properties of systems and are ubiquitously used for numerical calculations. A useful  class of spatio-temporal correlation functions are the \emph{Out-of-Time-Order Correlation functions} (OTOCs). 
They probe information transport properties of the induced dynamics, $U(0,t)$. 
Prominent examples of OTOCs are the two-point correlation function, four-point OTOCs, the $2k$-point OTOC, and most generally the $N$-time correlation function, of the form $\braket{O_1(t_1)O_2(t_2) \ldots O_n(t_n)}$, for some (local) operator $O_j$ and $O_j(t_j) := U(t_j,0)O_jU(0,t_j)$, with expectation value
is taken over all eigenstates of $U_t= U(0,t)$. 

The two-time correlation functions, $\braket{O_1(t)O_2}$ are commonly used in linear response theory \cite{Z65}, e.g., to study quantum quenches, describe oscillatory behaviour of dynamics on an ensemble of states, and in QCD calculations \cite{C16}, such as to compute the meson correlation function.
The four-point correlation functions, $\braket{O_1^{\dagger}(t)O_2^{\dagger}O_1(t)O_2}$, characterise transport properties of dynamical systems, such as information scrambling time and entanglement spreading velocities \cite{HLZMS19, BGL19}. The four-point OTOC has been used to numerically study phase transitions \cite{SCTHF20,DDS20,ZJL22} and quantum chaos \cite{MSS16,YK17,XS20}. 
The $2k$-point OTOCs have been shown to be 
a useful way to test if an ensemble forms a $k$-design \cite{RY17} i.e., forms a distribution approximately equal to the first $k$ moments of a Haar measure on the unitary group. More generally, the $N$-time correlation functions appear in calculations involving Green's functions and ubiquitously used for quantum field theoretic computations.

In our work, we investigate the precise extent of quantum computational resources required for estimating OTOCs, and, more generally, $N$-time correlation functions.
We study the problem of \emph{estimating correlation functions of local quantum dynamics on infinite temperature states}, where given classical description of local dynamics, e.g. as a sequence of one and two-qubit gates, resulting from Lie-Trotter-Suzuki or Magnus expansion, and local probes $\{O_j\}_j$, the computational problem is to estimate $\braket{O_1(t_1) \ldots O_N(t_N)}$ where the expectation is taken over infinite temperature, maximally mixed, states. When $t_1 \leq t_2 \leq \ldots \leq t_N$ the correlation function $\braket{O_1(t_1) \ldots O_N(t_N)}$ is called a N-Time (Ordered) Correlation function. Otherwise, when there is no partial order, the correlation function is called an Out-of-Time-Ordered Correlation function(OTOC).

Recent proposals to measure four-point OTOCs have been based on interferometry setups \cite{SBSH16, Yothers16, ZHG16}, 
and experimentally measured on synthetic materials using NMR simulators \cite{Lothers17}, 
Bose-Einstein Condensates \cite{MAG19}, 
superconducting systems~\cite{Bothers22}, 
and ion trap systems \cite{Lothers19} among others. These methods have been further extended to measure up to $10$-point correlation functions on NMR systems \cite{XPLSL17}. 
Measuring OTOCs with local operator probes however is an experimentally challenging task, partly due to the issues related to dissipation and time-reversal operation. Without the ability to faithfully implement the time-reversed dynamics, estimating properties of OTOCs can be intractable \cite{CSM23}. 

An alternative suggestion is to estimate OTOCs using ab initio methods by simulating the process on a quantum computer~\cite{SOGKL02, Pothers14, ARBP18, R20, Google21,PVNM21, DRFKF24} which circumvents the difficulties when running the corresponding experiments. Quantum algorithmic developments have led to an impressive suite of methods to estimate OTOCs, even in presence of noise~\cite{VESYZ19, YK17}, yielding useful results even in the absence of fault tolerance or scalable error mitigation techniques. 

\newpage
There have also been a number of new classical algorithms proposed for estimating OTOCs. These include numerical methods which track exact operator evolution by building an effective model of physical process~\cite{chen2017out}, taking advantage of matrix-product operators formalism to represent the evolution~\cite{xu2020accessing, bohrdt2017scrambling, hemery2019matrix} and even state-vector formalism using an exact time-evolution with a Krylov space method~\cite{luitz2017information}. Moreover, there have been several randomized algorithms to estimate these quantities: stochastic sampling methods~\cite{vidal2003efficient, zhou2023operator} or on matrix-product state approach coupled with the time-dependent variational principle~\cite{haegeman2016unifying, haegeman2011time}. These however have been limited in scale due to the inherent difficulty of simulating quantum dynamics on classical computers. 

Despite apparent computational hardness for classical methods for computing OTOCs for generic quantum systems, there exists only circumstantial evidence for the need for universal quantum computers \cite{LOZH21}. Given the rate of improvement of classical methods, it is suggestive that perhaps there might be no advantage in computing these quantities on a quantum computer at all.  However, this issue cannot be conclusively settled without a rigorous theoretical understanding of the complexity of estimating OTOCs. 

We prove here that the general problem of estimating OTOCs, and more generally, $n$-time correlation functions, for local dynamics over all eigenstates is DQC1-Complete; viz. OTOCs and $n$-time correlation functions can be estimated with the One Clean Qubit Model (often dubbed as DQC1~\cite{KL98}) and furthermore, estimating OTOCs is the hardest problem efficiently solvable in the DQC1 model. 

\begin{wrapfigure}{r}{0.3\textwidth}
\centering
\begin{quantikz}
\text{$\Ket{0}$}           &   \gate[2]{C}  & \meter{} \\
\text{$\frac{I}{2^n}$} \;  &  \qwbundle[alternate]{}                  & \qwbundle[alternate]{}  \\
\end{quantikz}
\caption{\label{DQC1} \small DQC1 computation with one clean and $n$ mixed state qubits}
\end{wrapfigure}
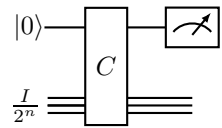

\textbf{One clean qubit model} (DQC1) is a computationally limited model of quantum computing using mixed states \cite{KL98}, with very low entanglement \cite{DFC05} and effective state space \cite{Z24}. The DQC1 model is conjectured to be strictly weaker than the class of universal polynomial-time quantum computation (BQP)~\cite{morimae2014hardness, ABKM17, JM24}.  
Despite these limitations, DQC1 is able to capture the complexity of a remarkable range of diverse problems such as trace estimation~\cite{KL98}, estimating Jones polynomials~\cite{SJ08}, estimating Schatten $p$-norms~\cite{CM18}, and determining whether the classical limit of a quantum system is chaotic or integrable~\cite{PLMP03}. 
From an experimental point of view, a DQC1 machine models idealised NMR systems with no decoherence or operational error.
One clean qubit model, illustrated in Fig \ref{DQC1}, is a model of mixed-state quantum computation on a system of $(1+n)$ qubits~\cite{DFC05} that consists of the following components: 
\begin{enumerate}
\item Initial state: $1+n$ qubits state, consisting of one qubit in pure state (`clean qubit') and mixed state on n-qubits; $\rho_I = \Ket{0}\Bra{0} \otimes \frac{I}{2^n}$.
\item  Local dynamics described as unitary circuit, $C_T = g_T \ldots g_3 g_2 g_1$, composed of $T = O(poly(n))$ many $SU(4)$ gates, $g_j$, acting on initial state $\rho_I$.
\item  Measurement:  The first qubit is measured in $\{\Ket{0}, \Ket{1}\}$ basis; 
$Prob(``0") = \Tr \Big[ \big(\Ket{0}\Bra{0}\otimes I\big) (C_T \rho_IC_T^{\dagger})  \Big]$
\end{enumerate}

\begin{definition}[OTOC] For any integer $k$, the class of $2k$-point out-of-time ordered correlation functions on $2^n$ dimensional Hilbert space, denoted as OTOC$_{\infty}$, can be defined as 
\begin{widetext}
\begin{align*}
	\nonumber
	OTOC_{\infty} & : U_t \times \{W_i\}_{i=1}^k \times \{V_i\}_{i=1}^k    \rightarrow  \mathbb{C}  \\	  	
		 OTOC_{\infty} & (U_t,W_1,V_1, \ldots W_k, V_k)   = \frac{1}{2^n}\Tr [W_1(t) V_1 \ldots W_k(t) V_k] = \langle W_1(t) V_1 \ldots W_k(t) V_k  \rangle,
\end{align*}
\noindent
\end{widetext}
where $W_i(t) = U_tW_iU_t^{\dagger}$ and $W_i, V_j$ are local operator matrices such that $[W_i,V_j] = 0$ for all $i,j \in [k]$,  and $U_t$ is a unitary matrix associated with (local) dynamics, that can be implemented using $poly(n)$ many $SU(4)$ matrices.
\end{definition}
\newpage

\textbf{DQC1-Hardness of OTOC$_{\infty}$}. We first establish the DQC1 hardness of estimating $2k$-point correlation functions over all eigenstates by showing the following:

{\bf Lemma 1}. Estimating OTOC$_{\infty}$ on $n$ qubits is DQC1-hard, for any non-negative integer $1\leq k \leq poly(n)$.

To show DQC1-hardness of estimating OTOC$_{\infty}$ over all eigenstates, it suffices to give a reduction $(\Phi, \phi)$ from a DQC1-hard problem, e.g. estimating the real part of normalised trace of unitary $C_T$ acting on $m$ qubits, 
$\widetilde{\Tr}:C_T \rightarrow  \frac{1}{2^m}Re\{\Tr[C_T]\} \in [-1,1]$ to the problem of estimating $2k$-point correlation function on $n=m+log_2(k)$ qubits, $OTOC_{\infty} \in [-1,1]$, such that the following diagram commutes,

\[
\begin{tikzcd}[column sep=5em, row sep=6em, every arrow/.append style={-latex, line width=1pt}
]
    \begin{tabular}{c} Unitary \\ $C_T$ \end{tabular} \arrow[r, "\Phi"] \arrow[d, "\widetilde{Tr}"] & 
    \begin{tabular}{c} \text{OTOC} \\ $\big\{U_t, \{W_j\}_j, \{V_j\}_j \big\}$ \end{tabular} \arrow[d, "OTOC_{\infty}"] \\
    \widetilde{Tr}[C_T] & \braket{W_1(t)V_1 \ldots W_k(t)V_k}  \arrow[l, "\phi"] 
\end{tikzcd}
\]

where $$ \Phi : C_T \rightarrow (U_t,W_1, V_1 \ldots W_k, V_k), \text{ s.t. } \widetilde{Tr}[C_T] = \phi \circ OTOC_{\infty} \circ \Phi[C_T] $$

Proof: For simplicity, let $k$ be any non-negative integer power of two. (It is easy to generalise this for $k$ which is not power of two, see Example~\ref{knotpowoftwo} in the Appendix.) The reduction makes use of a  generalised version of a gadget of~\cite{HORW}. Given a description of a circuit $C_T$, construct the image (output of the reduction map) of $\Phi$ as, 

\begin{enumerate}
    \item Set $U_t = \dmat{0}^{\otimes log_2(2k)} \otimes C_T + \sum_{j=1}^{2k-1} \dmat{j}\otimes I$. \\(When any $k \sim poly(n)$, this $U_t$ can be efficiently constructed  explicitly \cite{Bothers95}.)

    \item Set $W_j = \sigma_x^{(1)}, \forall j \in [k]$ (Pauli $\sigma_x$ acting on the first qubit)

    \item Set $V_j = \sigma_x^{(R(2j))}, \forall j \in [k-1]$, $V_k = \sigma_x^{(log_2(2k))}$   where $R(a)$ is the Ruler function that gives a 2-adic valuation of $2a$ and marks the position of the wire that changes at each step of the Gray code \cite{OEIS}. We write $\sigma_x^{(R(2j))}$ to denote Pauli $\sigma_x$ acting on the qubit indexed $R(2j)$.
\end{enumerate}

The map $\phi$ is simply the identity map, as $\phi(a)=a$.
One can verify that $\Phi[C_T] = (U_t,\{W_j\}, \{V_j\})$ is a valid instance of the OTOC$_{\infty}$, and and indeed $\phi(\braket{W_1(t)V_1 \ldots W_k(t) V_k}) = \widetilde{\Tr}[C_T]$ gives the exact equality, such that $\widetilde{Tr}[C_T] = \phi \circ OTOC_{\infty} \circ \Phi[C_T]$.    

This then shows that any DQC1 problem instance can be (Turing-) reduced to an instance of $2k$-point OTOC, and completes the proof of Lemma 1. $\hfill \square$

\textbf{DQC1 Membership of OTOC$_\infty$}. We now turn to describing how OTOC$_{\infty}$ can be efficiently estimated in DQC1.  We present an algorithm for estimating $2k$-point correlation functions, and show that OTOC$_{\infty}$ can be computed in DQC1. This construction follows by generalising previously suggested approaches~\cite{Pothers14, SBSH16, PVNM21}, which improves the circuit depth by a factor of two over previous approaches. 

{\bf Lemma 2}. Estimating $OTOC_{\infty}$ on $n$ qubits is in DQC1 for any $k,  1 \leq k \leq \in poly(n)$.

Proof: We first observe when $W_i,V_i$ are local operators acting on at most $l \sim O(1)$ qubits (let $l=2^\mathbb{N}$ for simplicity), $W_i,V_i$ 
 can be expressed in the Pauli basis (over complex field) as, 
$W_i = \sum_{w=1}^{4^l} \alpha_{w,i} \sigma_{w}$,  $V_i = \sum_{v=1}^{4^l} \alpha_{v,i} \sigma_{v}$; where each $\sigma_{j}$ is an element of the n-fold tensor product of Pauli matrices, and $\alpha_{w,i}, \alpha_{v,i} \in \mathbb{C}$.
Thus, without loss of generality, it suffices to only consider $W_i,V_i$ to be Pauli matrices, since estimating $2k$-point correlations can be expressed as,

\begin{equation*}
    \braket{W_1(t)V_1\ldots W_k(t)V_k} = \frac{1}{2^n}\sum_{w,v} \alpha_{w,1}\alpha_{v,1} \ldots \alpha_{w,k}\alpha_{v,k}     Tr[ \sigma_{w,1}(t) \sigma_{v,1} \ldots \sigma_{w,k}(t) \sigma_{v,k}] 
\end{equation*}
This readily follows from unique decomposition of $W,V$ into Pauli-basis and distributivity of matrix multiplication. Further, since $W_i, V_j$ are $l$-local matrices, the number of summands of the Pauli decomposition is bounded by $4^{2kl}$ summands and independent of system size $n$. 

Each term of the summand, $\braket{\sigma_{w,1}(t) \sigma_{v,1} \ldots \sigma_{w,k}(t) \sigma_{v,k} }$, can then be estimated using a DQC1-circuit as described in Figure \ref{2k_circ}, within $\epsilon > 0$ additive precision with probability of error at most $p$ using $O\big(log(1/p)/\epsilon^2\big)$ i.i.d. measurements of the given circuit. Using classical pre- and post-processing these summands can be added to faithfully estimate $\braket{W_1(t)V_1 \ldots W_k(t)V_k}$ up to $\varepsilon \sim \epsilon/2^{2k(l+1)}$ additive precision.


Finally, estimating $2k$-point correlation function is in DQC1, follows from the observation that the operator $\sigma_{w,1}(t) \sigma_{v,1} \ldots \sigma_{w,k}(t) \sigma_{v,k} =: \mathcal{O}$ is unitary, since the group of unitary matrices is closed under composition. 
Then estimating $OTOC_{\infty}$ is equivalent to estimating the trace of a unitary $\mathcal{O}$, which is in DQC1 \cite{SJ08}. More explicitly, $2k$ OTOC can be estimated by the circuit, as follows. 
Start with descriptions of $U$ acting on $n$ qubits, $2k$ probes $\{W_i\}_{i=1}^k$, $\{V_j\}_{j=1}^k$. Let $\Lambda_0(O) := \dmat{0}\otimes O + \dmat{1}\otimes I$, $\Lambda_1(O) := \dmat{0}\otimes I + \dmat{1}\otimes O$, $\widetilde{U} := I_2 \otimes U$ and $C \leftarrow I_N$. 
$C$ will contain the description of DQC1 circuit at the end of the routine. Then for $j=k/2:1$ times we perform the following operation: $C \leftarrow C \circ \big(\Lambda_1(V_{k-j+1}) \widetilde{U} \Lambda_1 (W_{k-j+1}) \Lambda_0(W_{j}) \widetilde{U}^{\dagger} \Lambda_0(V_{j}))$. The output is thus $C_T \leftarrow (H\otimes I_N) \circ C \circ (H \otimes I_N)$. 

The $2k$-point OTOC can then be faithfully estimated up to $\epsilon$ additive precision with $O(1/\epsilon^2)$ measurement samples of One Clean Qubit computation, estimating $Tr[(\dmat{0} \otimes I_N) C_T (\dmat{0} \otimes \frac{I_N}{N})C_T^{\dagger}]$.
Figure \ref{2k_circ} illustrates this DQC1-circuit which is used to obtain a measurement sample of $\braket{W_1(t)V_1 \ldots W_k(t)V_k}$. This then shows DQC1 membership of OTOC$_\infty$ as $2k$-point correlation functions over all eigenstates can be efficiently measured with one clean qubit model for any $1\leq k \leq poly(n)$. $\hfill \square$


\begin{widetext}
\begin{figure}[h]
\centering
\resizebox{0.7\textwidth}{!}{\begin{minipage}{\textwidth}
\begin{quantikz}
 \text{$\Ket{0}$} & \gate[1]{H}  & \octrl{1}      &\qw    & \octrl{3} & \ctrl{3} & \qw &  \ctrl{1}  &  \qw \ldots & \octrl{1}      &\qw    & \octrl{3} & \ctrl{3} & \qw &  \ctrl{1} & \gate[1]{H} & \meter{} \\
 \lstick[3]{$\frac{I}{2^n}$} & & \gate[1]{V_{\frac{k}{2}} }  & \gate[3]{U^{\dagger}}  & \qw  &\qw & \gate[3]{U} & \gate[1]{V_{\frac{k}{2}+1}}  & \qw \ldots\ & \gate[1]{V_1} & \gate[3]{U^{\dagger}}  & \qw  &\qw & \gate[3]{U} & \gate[1]{V_k} \\
  &  & \qwbundle[alternate]{}    & \qwbundle[alternate]{}         & \qwbundle[alternate]{} &    \qwbundle[alternate]{} & \qwbundle[alternate]{}  &   \qwbundle[alternate]{} & \ldots \qwbundle[alternate]{} &  \qwbundle[alternate]{}  & \qwbundle[alternate]{} & \qwbundle[alternate]{} & \qwbundle[alternate]{} & \qwbundle[alternate]{} & \qwbundle[alternate]{} \\  
  &          & \qw         &        &  \gate[1]{W_{\frac{k}{2}}}  & \gate[1]{W_{\frac{k}{2}+1}} & \qw & \qw & \qw \ldots & \qw         &        &  \gate[1]{W_1}  & \gate[1]{W_k} & \qw & \qw  \\
\end{quantikz}
\end{minipage}}
\caption{\label{2k_circ}
A circuit for $OTOC_{\infty}$ measuring normalised $2k$-point OTOC on $n$-qubit system, $\braket{W_1(t)V_1 \ldots W_k(t)V_k}$ as $Tr[(\dmat{0} \otimes I) C (\dmat{0} \otimes \frac{I}{2^n}) C^{\dagger} ]$}
\end{figure}
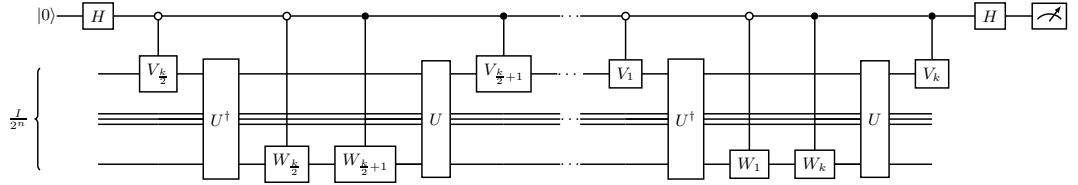
\end{widetext}

\noindent Having demonstrated the DQC1-hardness of OTOC$_\infty$ and DQC1-algorithm for estimating OTOC$_\infty$, we can state the main theorem that follows from Lemma 1 and Lemma 2,

\begin{theorem}\label{maintheorem}
Estimating $2k$-point correlation functions of local quantum dynamics over all eigenstates of a n-qubit system, OTOC$_\infty$, up to inverse poly(n) additive precision, is DQC1-Complete, for $1 \leq k \leq poly(n)$ $\square$
\end{theorem}

The DQC1-Completeness proof now leads to the following corollaries.

\begin{corollary}
\label{avgOTOC}
    Estimating \emph{average bipartite 4-point} OTOC on infinite temperature states on n qubits, up to $poly(n)$ precision, is DQC1-Complete, where average bipartite 4-point OTOC, with local probes, is $$\frac{1}{d^2_Wd^2_V}\sum_{W,V} \braket{W(t)VW(t)V} = \mathbb{E}_{W,V \sim \{I,X,Y,Z\}} \frac{1}{2^n}Tr[U_tWU_t^{\dagger}VU_tWU_t^{\dagger}V]$$ 
\end{corollary}

\noindent
The proof of Corollary follows by first giving a reduction map from DQC1-Complete class of problems to estimating average 4-point OTOC that establishes DQC1-hardness. We define the reduction map from any instance of trace estimation of $C_T = g_T \ldots g_2 g_1$ to an instance of averaged bipartite OTOC.
For simplicity, let the local subsystems associated with $W$ \& $V$ to be qubit 1 \& 2 resp, each of dimension $d_W = d_V = 2$. \\
Define $U_t = \dmat{0}_1 \otimes \dmat{0}_2 \otimes C_T + (I-\dmat{00}_{12})\otimes I_N = \prod_{\tau=1}^T \dmat{0}_1 \otimes \dmat{0}_2 \otimes g_{\tau} 
 + (I-\dmat{00}) \otimes I_N$,
$$ \frac{1}{16} \sum_{W,V \in \{I,X,Y,Z \}} \braket{W(t)VW(t)V} = \frac{3}{4} + \frac{4}{16} \widetilde{Tr}[C_T]$$ For instance when $W=X, V=Z$, then $\braket{W(t)VW(t)V}$ becomes $\braket{X(t)ZX(t)Z}$ with $X(t)=U_tXU_t^\dagger$. 

 DQC1 Membership of Bipartite average 4-point OTOC readily follows from the argument that each of the constant many summands $\braket{W(t)VW(t)V}$, corresponding to each of $W,V$ as l-fold Pauli operators, can be efficiently estimated on a DQC1-circuit. This then completes the proof of Corollary \ref{avgOTOC} $\square$

The bipartite averaged OTOC 
is exactly equal to the operator entanglement of a dynamical unitary \cite{SAZ21} and is a measure of the Reyni-2 entropy \cite{YK17}.  Alternate proposals to estimate these in the presence of noise has been proposed \cite{YK17,YY19} and demonstrated \cite{VESYZ19,Lothers19} in small scale devices. 

.

\begin{definition}[$N$-time correlation function] For any integer $N$, the class of $N$-time correlation functions on $2^n$ dimensional Hilbert space, denoted as $N-time_{\infty}$, can be defined as a function of a parameterised unitary $U(t)=U(0,t)$ acting on $n$-qubits and local operators $O_1 \ldots O_N$, 
as $$\text{N-time}_{\infty} : U(t) \times \{t_j \}_{j=1}^N \times \{O_j\}_{j=1}^N \rightarrow \braket{O_1(t_1) \ldots O_N(t_N) }.$$
\end{definition}

The problem of \noindent {\it Estimating $\text{N-time}$ correlation function} can then be stated as given as input a classical description of a parametrised unitary $U(t)$ acting on $n$ qubits, and $N$ time parameters ${t_1 \ldots t_N}$, and local operators $O_1, \ldots O_N$, estimate $\braket{O_1(t) \ldots O_N(t)}$ upto inverse polynomial additive precision,

\begin{corollary}
\label{TOC_DQC1}
    Estimating $N$-time correlation functions of local quantum dynamics over all eigenstates of a  m-qubit system, $TOC_{\infty}$ upto inverse poly(m) additive precision is DQC1-Complete. 
\end{corollary}

A $N$-time ordered correlation function is defined for a family of unitary matrices $U(t_j)$, and operators $O_j, O_j(t_j)= U(0,t_j)O_jU(t_j,0)$ as $\braket{O_1(t_1) \ldots O_N(t_N)}$, with $t_1 \leq \cdots \leq t_N$. This contrasts OTOCs where the unitary time evolution can be reversed, whereas for time ordered correlation functions, only forward evolution is allowed.  This DQC1-Completeness proof of the $2k$-point correlation functions, \cref{maintheorem}, directly extends to also show that the $N$-time correlation function acting on $m$-qubits, $\braket{\mathcal{O}_1(t_1)\mathcal{O}_2(t_2) \ldots \mathcal{O}_N(t_N)}$ is DQC1-Complete for $t_N \in O(poly(n))$.

Theorem~\ref{maintheorem} and its corollaries lead to a range of interesting applications.

\textbf{Probing hydrodynamical transport in spin systems.}
Simulating large scale quantum mechanical systems at infinite temperature, to probe transport properties remain of significant interest in the the study of equilibrium and non-equilibrium physics. The diffusive or ballistic growth of operators offer insights to the fundamental study of exotic phases of matter. Thus far, they are probed experimentally such as in analog quantum devices e.g., using autocorrelations functions of the form $\braket{O(t)O}$, such as done in \cite{shi2023probing}. The DQC1-Completeness of OTOCs now indicate that these processes can be simulated on a conjecturally weaker models of models.

{\bf Probing ground state phase transitions in DQC1.} 
A useful application of the four-point OTOCs has been probing quantum phase transitions. 
The time-averaged of infinite-temperature OTOCs, defined as $\frac{1}{\mathcal{T}} \int \braket{W(t)VW(t)V} dt$, for some time interval $\mathcal{T}$, can probe the zero-temperature quantum phases, for generic dynamical systems that can be mapped to a 1D Majorana chain (such as the XXZ model), by detecting the presence of Majorana zero modes at the ends of 1D-chain with $Z_2$ topological order \cite{DDS20}.
The DQC1-membership of infinite temperature OTOCs then immediately shows that probing dynamical properties of (bulk) ground states of generic (non-)interacting, non-integrable models that can be mapped to a 1D Majorana chain (e.g. via Jordan Wigner or similar transformations) is in DQC1. 
This indicates that the computational complexity of task maybe substantially lower than previously expected.

{\bf Discussions}  
The DQC1 hardness of OTOCs and $N$-time correlation functions extends the findings reported in \cite{PVNM21}, where it was argued that a special class of (four-point) OTOCs can be estimated using the DQC1 model, but leaving open the question about the computation complexity of estimating OTOCs. Surprisingly, it was shown in \cite{Google21} that for a broad class of random circuit ensembles, mean value of the OTOC (averaged over the ensemble itself) can be in fact estimated efficiently, classically, using Markov population processes. It was further claimed in \cite{Google21} that estimating OTOC fluctuations (root-mean-squared values of OTOC over the ensemble) was computationally challenging. Our results establishing DQC1-hardness of OTOCs then resolves the question about the computational complexity of estimating OTOCs by classifying the hardness.

This characterisation of hardness of $2k$-point OTOCs then also complements the results obtained in \cite{LOZH21}, where authors  argue that the maximal decay of higher order Out-of-Time-Order correlations necessarily requires non-Clifford (universal) resources and may not be classically simulated, in the sense that our results show that (conjecturally) weaker model of computation, viz. DQC1, with a significantly smaller effective state space, is sufficient for observing the maximal decay of higher order OTOCs, neither need the full power of universal quantum computers, nor quantum theory which otherwise has a larger effective state space.

Lastly, the DQC1-Completeness of estimating infinite temperature $2K$-point Out-of-Time-Ordered correlation functions (Theorem~\ref{maintheorem}) together with the DQC1-Completeness of estimating infinite temperature $N$-time correlation function (Corollary~\ref{TOC_DQC1}) show them to be computationally equivalent. This can be contrasted with the results obtained in \cite{CSM23} that show given only query access to a unitary (and without access to the time-reversed unitary) there exists an exponential separation in query complexity, for estimating OTOCs and N-time correlation functions.

\textbf{Acknowledgements}.
We thank Benjamin Beri for comments on this manuscript. SS acknowledges support from the Royal Society University Research Fellowship. SRM and SS acknowledge ``Quantum simulation algorithms for quantum chromodynamics'' grant (ST/W006251/1).

\bibliographystyle{unsrt}
\bibliography{bibliography}

\begin{thebibliography}{10}

\bibitem{Z65}
Robert Zwanzig.
\newblock Time-correlation functions and transport coefficients in statistical
  mechanics.
\newblock {\em Annual Review of Physical Chemistry}, 16(1):67--102, 1965.

\bibitem{C16}
Bipasha Chakraborty.
\newblock {\em Precision tests of the Standard Model using lattice QCD}.
\newblock PhD thesis, University of Glasgow, 2016.

\bibitem{HLZMS19}
Aram~W Harrow, Linghang Kong, Zi-Wen Liu, Saeed Mehraban, and Peter~W Shor.
\newblock Separation of out-of-time-ordered correlation and entanglement.
\newblock {\em PRX Quantum}, 2(2):020339, 2021.

\bibitem{BGL19}
Gregory Bentsen, Yingfei Gu, and Andrew Lucas.
\newblock Fast scrambling on sparse graphs.
\newblock {\em Proceedings of the National Academy of Sciences},
  116(14):6689--6694, 2019.

\bibitem{SCTHF20}
Zheng-Hang Sun, Jia-Qi Cai, Qi-Cheng Tang, Yong Hu, and Heng Fan.
\newblock Out-of-time-order correlators and quantum phase transitions in the
  rabi and dicke models.
\newblock {\em Annalen der Physik}, 532(4):1900270, 2020.

\bibitem{DDS20}
Ceren~B Da{\u{g}}, L-M Duan, and Kai Sun.
\newblock Topologically induced prescrambling and dynamical detection of
  topological phase transitions at infinite temperature.
\newblock {\em Physical Review B}, 101(10):104415, 2020.

\bibitem{ZJL22}
Sara Zamani, R~Jafari, and A~Langari.
\newblock Out-of-time-order correlations and floquet dynamical quantum phase
  transition.
\newblock {\em Physical Review B}, 105(9):094304, 2022.

\bibitem{MSS16}
Juan Maldacena, Stephen~H Shenker, and Douglas Stanford.
\newblock A bound on chaos.
\newblock {\em Journal of High Energy Physics}, 2016(8):1--17, 2016.

\bibitem{YK17}
Beni Yoshida and Alexei Kitaev.
\newblock Efficient decoding for the hayden-preskill protocol.
\newblock {\em arXiv:1710.03363}, 2017.

\bibitem{XS20}
Shenglong Xu and Brian Swingle.
\newblock Accessing scrambling using matrix product operators.
\newblock {\em Nature Physics}, 16(2):199--204, 2020.

\bibitem{RY17}
Daniel~A Roberts and Beni Yoshida.
\newblock Chaos and complexity by design.
\newblock {\em Journal of High Energy Physics}, 2017(4):1--64, 2017.

\bibitem{SBSH16}
Brian Swingle, Gregory Bentsen, Monika Schleier-Smith, and Patrick Hayden.
\newblock Measuring the scrambling of quantum information.
\newblock {\em Physical Review A}, 94(4):040302, 2016.

\bibitem{Yothers16}
Norman~Y Yao, Fabian Grusdt, Brian Swingle, Mikhail~D Lukin, Dan~M
  Stamper-Kurn, Joel~E Moore, and Eugene~A Demler.
\newblock Interferometric approach to probing fast scrambling.
\newblock {\em arXiv:1607.01801}, 2016.

\bibitem{ZHG16}
Guanyu Zhu, Mohammad Hafezi, and Tarun Grover.
\newblock Measurement of many-body chaos using a quantum clock.
\newblock {\em Physical Review A}, 94(6):062329, 2016.

\bibitem{Lothers17}
Jun Li, Ruihua Fan, Hengyan Wang, Bingtian Ye, Bei Zeng, Hui Zhai, Xinhua Peng,
  and Jiangfeng Du.
\newblock Measuring out-of-time-order correlators on a nuclear magnetic
  resonance quantum simulator.
\newblock {\em Physical Review X}, 7(3):031011, 2017.

\bibitem{MAG19}
Eric~J Meier, Fangzhao~Alex An, and Bryce Gadway.
\newblock Exploring quantum signatures of chaos on a floquet synthetic lattice.
\newblock {\em Physical Review A}, 100(1):013623, 2019.

\bibitem{Bothers22}
Jochen Braum{\"u}ller, Amir~H Karamlou, Yariv Yanay, Bharath Kannan, David Kim,
  Morten Kjaergaard, Alexander Melville, Bethany~M Niedzielski, Youngkyu Sung,
  Antti Veps{\"a}l{\"a}inen, et~al.
\newblock Probing quantum information propagation with out-of-time-ordered
  correlators.
\newblock {\em Nature Physics}, 18(2):172--178, 2022.

\bibitem{Lothers19}
Kevin~A Landsman, Caroline Figgatt, Thomas Schuster, Norbert~M Linke, Beni
  Yoshida, Norman~Y Yao, and Christopher Monroe.
\newblock Verified quantum information scrambling.
\newblock {\em Nature}, 567(7746):61--65, 2019.

\bibitem{XPLSL17}
Tao Xin, Julen~S Pedernales, Lucas Lamata, Enrique Solano, and Gui-Lu Long.
\newblock Measurement of linear response functions in nuclear magnetic
  resonance.
\newblock {\em Scientific Reports}, 7(1):12797, 2017.

\bibitem{CSM23}
Jordan Cotler, Thomas Schuster, and Masoud Mohseni.
\newblock Information-theoretic hardness of out-of-time-order correlators.
\newblock {\em Physical Review A}, 108(6):062608, 2023.

\bibitem{SOGKL02}
Rolando Somma, Gerardo Ortiz, James~E Gubernatis, Emanuel Knill, and Raymond
  Laflamme.
\newblock Simulating physical phenomena by quantum networks.
\newblock {\em Physical Review A}, 65(4):042323, 2002.

\bibitem{Pothers14}
JS~Pedernales, R~Di~Candia, IL~Egusquiza, J~Casanova, and Enrique Solano.
\newblock Efficient quantum algorithm for computing n-time correlation
  functions.
\newblock {\em Physical Review Letters}, 113(2):020505, 2014.

\bibitem{ARBP18}
Daattavya Aggarwal, Shivam Raj, Bikash~K Behera, and Prasanta~K Panigrahi.
\newblock Application of quantum scrambling in rydberg atom on ibm quantum
  computer.
\newblock {\em arXiv:1806.00781}, 2018.

\bibitem{R20}
Patrick Rall.
\newblock Quantum algorithms for estimating physical quantities using block
  encodings.
\newblock {\em Physical Review A}, 102(2):022408, 2020.

\bibitem{Google21}
Xiao Mi, Pedram Roushan, Chris Quintana, Salvatore Mandra, Jeffrey Marshall,
  Charles Neill, Frank Arute, Kunal Arya, Juan Atalaya, Ryan Babbush, et~al.
\newblock Information scrambling in computationally complex quantum circuits.
\newblock {\em arXiv:2101.08870}, 2021.

\bibitem{PVNM21}
Sreeram Pg, Naga~Dileep Varikuti, and Vaibhav Madhok.
\newblock Exponential speedup in measuring out-of-time-ordered correlators and
  gate fidelity with a single bit of quantum information.
\newblock {\em Physics Letters A}, 397:127257, 2021.

\bibitem{DRFKF24}
Lorenzo Del~Re, Brian Rost, Michael Foss-Feig, AF~Kemper, and JK~Freericks.
\newblock Robust measurements of n-point correlation functions of
  driven-dissipative quantum systems on a digital quantum computer.
\newblock {\em Physical Review Letters}, 132(10):100601, 2024.

\bibitem{VESYZ19}
Beno{\^\i}t Vermersch, Andreas Elben, Lukas~M Sieberer, Norman~Y Yao, and Peter
  Zoller.
\newblock Probing scrambling using statistical correlations between randomized
  measurements.
\newblock {\em Physical Review X}, 9(2):021061, 2019.

\bibitem{chen2017out}
Xiao Chen, Tianci Zhou, David~A Huse, and Eduardo Fradkin.
\newblock Out-of-time-order correlations in many-body localized and thermal
  phases.
\newblock {\em Annalen der Physik}, 529(7):1600332, 2017.

\bibitem{xu2020accessing}
Shenglong Xu and Brian Swingle.
\newblock Accessing scrambling using matrix product operators.
\newblock {\em Nature Physics}, 16(2):199--204, 2020.

\bibitem{bohrdt2017scrambling}
Annabelle Bohrdt, Christian~B Mendl, Manuel Endres, and Michael Knap.
\newblock Scrambling and thermalization in a diffusive quantum many-body
  system.
\newblock {\em New Journal of Physics}, 19(6):063001, 2017.

\bibitem{hemery2019matrix}
K{\'e}vin H{\'e}mery, Frank Pollmann, and David~J Luitz.
\newblock Matrix product states approaches to operator spreading in ergodic
  quantum systems.
\newblock {\em Physical Review B}, 100(10):104303, 2019.

\bibitem{luitz2017information}
David~J Luitz and Yevgeny~Bar Lev.
\newblock Information propagation in isolated quantum systems.
\newblock {\em Physical Review B}, 96(2):020406, 2017.

\bibitem{vidal2003efficient}
Guifr{\'e} Vidal.
\newblock Efficient classical simulation of slightly entangled quantum
  computations.
\newblock {\em Physical review letters}, 91(14):147902, 2003.

\bibitem{zhou2023operator}
Tianci Zhou and Brian Swingle.
\newblock Operator growth from global out-of-time-order correlators.
\newblock {\em Nature communications}, 14(1):3411, 2023.

\bibitem{haegeman2016unifying}
Jutho Haegeman, Christian Lubich, Ivan Oseledets, Bart Vandereycken, and Frank
  Verstraete.
\newblock Unifying time evolution and optimization with matrix product states.
\newblock {\em Physical Review B}, 94(16):165116, 2016.

\bibitem{haegeman2011time}
Jutho Haegeman, J~Ignacio Cirac, Tobias~J Osborne, Iztok Pi{\v{z}}orn, Henri
  Verschelde, and Frank Verstraete.
\newblock Time-dependent variational principle for quantum lattices.
\newblock {\em Physical review letters}, 107(7):070601, 2011.

\bibitem{LOZH21}
Lorenzo Leone, Salvatore~FE Oliviero, You Zhou, and Alioscia Hamma.
\newblock Quantum chaos is quantum.
\newblock {\em Quantum}, 5:453, 2021.

\bibitem{KL98}
Emanuel Knill and Raymond Laflamme.
\newblock Power of one bit of quantum information.
\newblock {\em Physical Review Letters}, 81(25):5672, 1998.

\bibitem{DFC05}
Animesh Datta, Steven~T Flammia, and Carlton~M Caves.
\newblock Entanglement and the power of one qubit.
\newblock {\em Physical Review A}, 72(4):042316, 2005.

\bibitem{Z24}
Zachary Stier.
\newblock A no-go result for pure state synthesis in the dqc1 model.
\newblock {\em arXiv preprint arXiv:2404.04198}, 2024.

\bibitem{morimae2014hardness}
Tomoyuki Morimae, Keisuke Fujii, and Joseph~F Fitzsimons.
\newblock Hardness of classically simulating the one-clean-qubit model.
\newblock {\em Physical review letters}, 112(13):130502, 2014.

\bibitem{ABKM17}
Scott Aaronson, Adam Bouland, Greg Kuperberg, and Saeed Mehraban.
\newblock The computational complexity of ball permutations.
\newblock In {\em Proceedings of the 49th Annual ACM SIGACT Symposium on Theory
  of Computing}, pages 317--327, 2017.

\bibitem{JM24}
Dale Jacobs and Saeed Mehraban.
\newblock The space just above one clean qubit.
\newblock {\em arXiv preprint arXiv:2410.08051}, 2024.

\bibitem{SJ08}
Peter~W Shor and Stephen~P Jordan.
\newblock Estimating jones polynomials is a complete problem for one clean
  qubit.
\newblock {\em Quantum Information and Computation}, 8(8):681–714, 2008.

\bibitem{CM18}
Chris Cade and Ashley Montanaro.
\newblock The quantum complexity of computing schatten $p$-norms.
\newblock In {\em 45th International Colloquium on Automata, Languages, and
  Programming (ICALP 2018)}, pages 10:1--10:14. Schloss
  Dagstuhl--Leibniz-Zentrum fuer Informatik, 2018.

\bibitem{PLMP03}
David Poulin, Raymond Laflamme, GJ~Milburn, and Juan~Pablo Paz.
\newblock Testing integrability with a single bit of quantum information.
\newblock {\em Physical Review A}, 68(2):022302, 2003.

\bibitem{HORW}
Havlicek, O'Gorman, Roy Moulik, and Wojcan.
\newblock Computational hardness of out-of-time-order correlation functions on
  pure states.
\newblock {\em unpublished}.

\bibitem{Bothers95}
Adriano Barenco, Charles~H Bennett, Richard Cleve, David~P DiVincenzo, Norman
  Margolus, Peter Shor, Tycho Sleator, John~A Smolin, and Harald Weinfurter.
\newblock Elementary gates for quantum computation.
\newblock {\em Physical review A}, 52(5):3457, 1995.

\bibitem{OEIS}
OEIS Foundation.
\newblock The on-line encyclopedia of integer sequences, {A}001511.
\newblock Accessed on Dec 15, 2023.

\bibitem{SAZ21}
Georgios Styliaris, Namit Anand, and Paolo Zanardi.
\newblock Information scrambling over bipartitions: Equilibration, entropy
  production, and typicality.
\newblock {\em Physical Review Letters}, 126(3):030601, 2021.

\bibitem{YY19}
Beni Yoshida and Norman~Y Yao.
\newblock Disentangling scrambling and decoherence via quantum teleportation.
\newblock {\em Physical Review X}, 9(1):011006, 2019.

\bibitem{shi2023probing}
Yun-Hao Shi, Zheng-Hang Sun, Yong-Yi Wang, Zheng-An Wang, Yu-Ran Zhang, Wei-Guo
  Ma, Hao-Tian Liu, Kui Zhao, Jia-Cheng Song, Gui-Han Liang, et~al.
\newblock Probing spin hydrodynamics on a superconducting quantum simulator.
\newblock {\em arXiv preprint arXiv:2310.06565}, 2023.

\end{thebibliography}

\newpage

\appendix

\section{Supplemental Information}

We give examples of reduction maps that show DQC1 Hardness of 2, 4, 6, 8, and 16 point OTOCs.

\noindent
First recall the reduction map \cite{KL98}, that show DQC1 hardness of normalised trace estimation of a $poly(m)$-sized circuit acting on $m+1$ qubits. For all the subsequent examples we consider $C$ to be a $poly(m)$-sized quantum circuit on $m$ qubits.

\begin{example}
     The reduction $\Phi_2$ maps an instance of a DQC1-Complete problem, of estimating the normalised trace of a circuit $C$ on $m$ qubits, $\widetilde\Tr[C]$, to an instance of an estimating 2-point OTOC, $\braket{UWU^{\dagger}V}$ on $n=m+1$ qubits, as $\Phi_2[C] = (U, W, V)$, with
$U = \dmat{0}\otimes C + \dmat{1}\otimes \mathds{I}_{2^m}$, $W = V = \sigma_x^{(1)}$. Then it follows, $Re[\widetilde\Tr[C]] \propto OTOC(\Phi_2[C]) = \braket{W(t)V} = \frac{1}{2^n}Tr[UWU^{\dagger}V]$. This completes the sketch of the reduction map and shows DQC1-hardness of estimating $2$-point OTOCs.

\end{example}

\begin{example}
     The reduction $\Phi_4$ similarly maps an instance of a DQC1-Complete problem, of estimating the normalised trace of a circuit $C$ on $m$ qubits, $\widetilde\Tr[C]$, to an instance of an estimating 4-point OTOC, $\braket{UWU^{\dagger}VUWU^{\dagger}V}$ on $n=m+2$ qubits, as $\Phi_2[C] = (U, W, V)$, with
$U = \dmat{0}\otimes C + (\mathds{I}_{2^2}- \dmat{0})\otimes \mathds{I}_{2^m}$, 
$W = V = \sigma_x^{(1)}$. It then follows, $Re[\widetilde\Tr[C]] \propto OTOC(\Phi_4[C]) = \braket{W(t)VW(t)V} = \frac{1}{2^n}Tr[UWU^{\dagger}V]$ and completes the sketch of DQC1-Hardness of estimating 4-point OTOCs.

\begin{widetext}
\begin{figure}[h]
\centering
\resizebox{\textwidth}{!}{\begin{minipage}{\textwidth}

\begin{quantikz}[transparent, row sep={0.8cm,between origins}]
\qw & \ctrl{0} & \gate{X} & \ctrl{0} & \qw & \ctrl{0} & \gate{X} & \ctrl{0} & \qw & \qw\\
\qw & \ctrl{-1} & \qw & \ctrl{-1} & \gate{X} & \ctrl{-1} & \qw & \ctrl{-1} & \gate{X} & \qw\\
\qw & \gate[2,disable auto height]{C}\vqw{-1} & \qw & \gate[2,disable auto height]{C^+}\vqw{-1} & \qw & \gate[2,disable auto height]{C}\vqw{-1} & \qw & \gate[2,disable auto height]{C^+}\vqw{-1} & \qw & \qw\\
\qw & \qw & \qw & \qw & \qw & \qw & \qw & \qw & \qw & \qw
\end{quantikz}

\end{minipage}}
\caption{\label{dqc1_circ} Schematic of the $4$-point OTOC circuit described by $\Phi_4[C]$ that estimates normalised trace of circuit $C$.}
\end{figure}
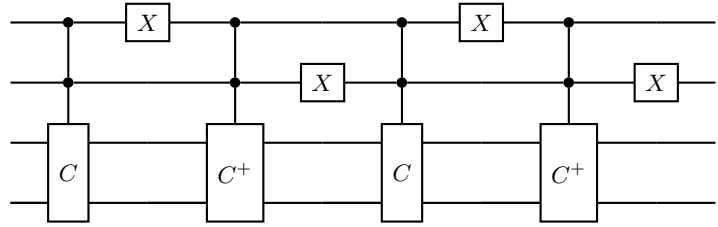
\end{widetext}

\end{example}

\begin{example}
 The reduction $\Phi_6$ maps an instance of a DQC1-Complete problem, of estimating   $\widetilde\Tr[C]$, to an instance of an 6-point OTOC, $\braket{UW_1U^{\dagger}V_1UW_2U^{\dagger}V_2UW_3U^{\dagger}V_3}$ on $n=m+3$ qubits, as $\Phi_6[C] = (U, W_1\ldots W_3, V_1, \ldots, V_3)$, with
$U = \dmat{000}\otimes C + (\mathds{I}_{2^3}- \dmat{000})\otimes \mathds{I}_{2^m}$, \\
$W_1 = V_2 = \sigma_x^{(1)}, 
 V_1 = W_3 = \sigma_x^{(2)} ,
 W_2 = V_3 = \sigma_x^{(3)}$. Then, $Re[\widetilde\Tr[C]] \propto OTOC(\Phi_6[C]) = \braket{W_1(t)V_1W_2(t)V_2 W_3(t)VW}$. This shows DQC1-hardness  of estimating 6-point OTOCs.

\begin{widetext}
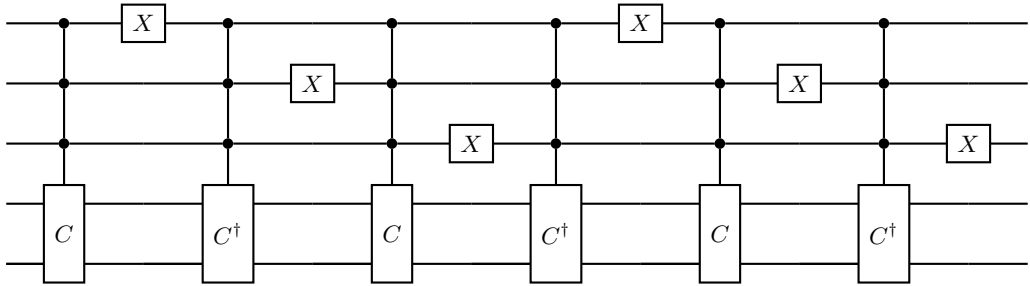
\begin{figure}[h]
\centering
\resizebox{\textwidth}{!}{\begin{minipage}{\textwidth}

\begin{quantikz}[transparent, row sep={0.8cm,between origins}]
\qw & \ctrl{0} & \gate{X} & \ctrl{0} & \qw & \ctrl{0} & \qw & \ctrl{0} & \gate{X} & \ctrl{0} & \qw & \ctrl{0} & \qw & \qw\\
\qw & \ctrl{-1} & \qw & \ctrl{-1} & \gate{X} & \ctrl{-1} & \qw & \ctrl{-1} & \qw & \ctrl{-1} & \gate{X} & \ctrl{-1} & \qw & \qw\\
\qw & \ctrl{-1} & \qw & \ctrl{-1} & \qw & \ctrl{-1} & \gate{X} & \ctrl{-1} & \qw & \ctrl{-1} & \qw & \ctrl{-1} & \gate{X} & \qw\\
\qw & \gate[2,disable auto height]{C}\vqw{-1} & \qw & \gate[2,disable auto height]{C^{\dagger}}\vqw{-1} & \qw & \gate[2,disable auto height]{C}\vqw{-1} & \qw & \gate[2,disable auto height]{C^{\dagger}}\vqw{-1} & \qw & \gate[2,disable auto height]{C}\vqw{-1} & \qw & \gate[2,disable auto height]{C^{\dagger}}\vqw{-1} & \qw & \qw\\
\qw & \qw & \qw & \qw & \qw & \qw & \qw & \qw & \qw & \qw & \qw & \qw & \qw & \qw
\end{quantikz}

\end{minipage}}
\caption{\label{dqc1_circ} Schematic of the $6$-point OTOC circuit described by $\Phi_6[C]$ that estimates normalised trace of circuit $C$.}
\end{figure}
\end{widetext}

\end{example}

\begin{example}
\label{knotpowoftwo}  The reduction, $\Phi_8$, maps an instance of a DQC1-Complete problem of estimating the normalised trace of $C$ on $m$ qubits, $\widetilde\Tr[C]$, to 8-point OTOC on $n=m+3$ qubits.  
The 8-point OTOC is defined as $\braket{UW_1U^{\dagger}V_1UW_2U^{\dagger}V_2UW_3U^{\dagger}V_3UW_4U^{\dagger}V_4}$ and with $\Phi_8[C] = (U, W_1\ldots W_3, V_1, \ldots, V_4)$, as
$U = \dmat{000}\otimes C + (\mathds{I}_{2^3}- \dmat{000})\otimes \mathds{I}_{2^m}$, $W_1 = W_2 = W_3 = W_4 = \sigma_x^{(1)}$, 
$V_1 = \sigma_x^{(2)} ,
 V_2 = \sigma_x^{(3)} ,
 V_3 = \sigma_x^{(2)} ,
 V_4=  \sigma_x^{(3)} $
\noindent The 8-point OTOC instance, $\Phi_8[C]$, can be summarised by the following diagrammatic calculus, and verified that
$$OTOC(\Phi_8[C]) = \widetilde\Tr[W_1(t)V_1W_2(t)V_2W_3(t)V_3W_4(t)V_4] = \frac{1}{2}(\widetilde\Tr[C]+ \widetilde\Tr[C^{\dagger}]) =  Re\big[\widetilde\Tr[C]\big]$$.

\begin{widetext}
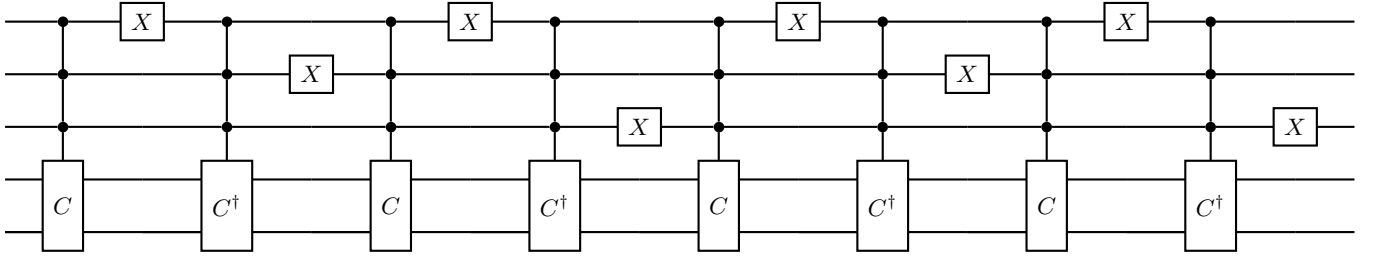
\begin{figure}[h]
\centering
\resizebox{\textwidth}{!}{\begin{minipage}{\textwidth}

\begin{quantikz}[transparent, row sep={0.7cm,between origins}]
\qw & \ctrl{0} & \gate{X} & \ctrl{0} & \qw & \ctrl{0} & \gate{X} & \ctrl{0} & \qw & \ctrl{0} & \gate{X} & \ctrl{0} & \qw & \ctrl{0} & \gate{X} & \ctrl{0} & \qw & \qw\\
\qw & \ctrl{-1} & \qw & \ctrl{-1} & \gate{X} & \ctrl{-1} & \qw & \ctrl{-1} & \qw & \ctrl{-1} & \qw & \ctrl{-1} & \gate{X} & \ctrl{-1} & \qw & \ctrl{-1} & \qw & \qw\\
\qw & \ctrl{-1} & \qw & \ctrl{-1} & \qw & \ctrl{-1} & \qw & \ctrl{-1} & \gate{X} & \ctrl{-1} & \qw & \ctrl{-1} & \qw & \ctrl{-1} & \qw & \ctrl{-1} & \gate{X} & \qw\\
\qw & \gate[2,disable auto height]{C}\vqw{-1} & \qw & \gate[2,disable auto height]{C^{\dagger}}\vqw{-1} & \qw & \gate[2,disable auto height]{C}\vqw{-1} & \qw & \gate[2,disable auto height]{C^{\dagger}}\vqw{-1} & \qw & \gate[2,disable auto height]{C}\vqw{-1} & \qw & \gate[2,disable auto height]{C^{\dagger}}\vqw{-1} & \qw & \gate[2,disable auto height]{C}\vqw{-1} & \qw & \gate[2,disable auto height]{C^{\dagger}}\vqw{-1} & \qw & \qw\\
\qw & \qw & \qw & \qw & \qw & \qw & \qw & \qw & \qw & \qw & \qw & \qw & \qw & \qw & \qw & \qw & \qw & \qw
\end{quantikz}

\end{minipage}}
\caption{\label{dqc1_circ} 
Schematic of the $8$-point OTOC circuit described by $\Phi_8[C]$ that estimates normalised trace of circuit $C$.}
\end{figure}
\end{widetext}
\end{example}

\begin{example} The reduction, $\Phi_{16}$, maps an instance of a DQC1-Complete problem of estimating normalised trace of a circuit $C$ on $m$ qubits, $\widetilde\Tr[C]$, to estimating 16-point OTOC acting on $n=m+4$ qubits. The reduction map to the OTOC,  $\braket{UW_1U^{\dagger}V_1UW_2U^{\dagger}V_2 \ldots UW_8U^{\dagger}V_8}$,  is given as $\Phi_8[C] = (U, W_1\ldots W_8, V_1, \ldots, V_8)$, with,
$U = \dmat{0000}\otimes C + (\mathds{I}_{2^4}- \dmat{0000})\otimes \mathds{I}_{2^m}$, 
$W_j = \sigma_x^{(1)}$, for all $j \in \{1 \ldots 8\}$;  and 
$V_1 = \sigma_x^{(2)}$ ,
 $V_2 = \sigma_x^{(3)}$ ,
 $V_3 = \sigma_x^{(2)}$ ,
 $V_4=  \sigma_x^{(4)}$, 
 $V_5 = \sigma_x^{(2)}$ ,
$ V_6 = \sigma_x^{(3)} $,
$ V_7 = \sigma_x^{(2)} $,
$ V_8 = \sigma_x^{(4)}.$

The 16-point Correlation function, 
$OTOC(\Phi_{16}[C]) = \widetilde\Tr[W_1(t)V_1W_2(t)V_2W_3(t)V_3W_4(t)V_4] = Re\big[\widetilde\Tr[C]\big]$. The OTOC instance, $\Phi_8[C]$, can be summarised by the following diagrammatic calculus,

\begin{figure}[h]
\scalebox{0.5}{
\begin{quantikz}[transparent, row sep={1cm,between origins}]  
\qw & \ctrl{0} & \gate{X} & \ctrl{0} & \qw & \ctrl{0} & \gate{X} & \ctrl{0} & \qw & \ctrl{0} & \gate{X} & \ctrl{0} & \qw & \ctrl{0} & \gate{X} & \ctrl{0} & \qw & \ctrl{0} & \gate{X} & \ctrl{0} & \qw & \ctrl{0} & \gate{X} & \ctrl{0} & \qw & \ctrl{0} & \gate{X} & \ctrl{0} & \qw & \ctrl{0} & \gate{X} & \ctrl{0} & \qw & \qw\\
\qw & \ctrl{-1} & \qw & \ctrl{-1} & \gate{X} & \ctrl{-1} & \qw & \ctrl{-1} & \qw & \ctrl{-1} & \qw & \ctrl{-1} & \gate{X} & \ctrl{-1} & \qw & \ctrl{-1} & \qw & \ctrl{-1} & \qw & \ctrl{-1} & \gate{X} & \ctrl{-1} & \qw & \ctrl{-1} & \qw & \ctrl{-1} & \qw & \ctrl{-1} & \gate{X} & \ctrl{-1} & \qw & \ctrl{-1} & \qw & \qw\\
\qw & \ctrl{-1} & \qw & \ctrl{-1} & \qw & \ctrl{-1} & \qw & \ctrl{-1} & \gate{X} & \ctrl{-1} & \qw & \ctrl{-1} & \qw & \ctrl{-1} & \qw & \ctrl{-1} & \qw & \ctrl{-1} & \qw & \ctrl{-1} & \qw & \ctrl{-1} & \qw & \ctrl{-1} & \gate{X} & \ctrl{-1} & \qw & \ctrl{-1} & \qw & \ctrl{-1} & \qw & \ctrl{-1} & \qw & \qw\\
\qw & \ctrl{-1} & \qw & \ctrl{-1} & \qw & \ctrl{-1} & \qw & \ctrl{-1} & \qw & \ctrl{-1} & \qw & \ctrl{-1} & \qw & \ctrl{-1} & \qw & \ctrl{-1} & \gate{X} & \ctrl{-1} & \qw & \ctrl{-1} & \qw & \ctrl{-1} & \qw & \ctrl{-1} & \qw & \ctrl{-1} & \qw & \ctrl{-1} & \qw & \ctrl{-1} & \qw & \ctrl{-1} & \gate{X} & \qw\\
\qw & \gate[2,disable auto height]{C}\vqw{-1} & \qw & \gate[2,disable auto height]{C^{\dagger}}\vqw{-1} & \qw & \gate[2,disable auto height]{C}\vqw{-1} & \qw & \gate[2,disable auto height]{C^{\dagger}}\vqw{-1} & \qw & \gate[2,disable auto height]{C}\vqw{-1} & \qw & \gate[2,disable auto height]{C^{\dagger}}\vqw{-1} & \qw & \gate[2,disable auto height]{C}\vqw{-1} & \qw & \gate[2,disable auto height]{C^{\dagger}}\vqw{-1} & \qw & \gate[2,disable auto height]{C}\vqw{-1} & \qw & \gate[2,disable auto height]{C^{\dagger}}\vqw{-1} & \qw & \gate[2,disable auto height]{C}\vqw{-1} & \qw & \gate[2,disable auto height]{C^{\dagger}}\vqw{-1} & \qw & \gate[2,disable auto height]{C}\vqw{-1} & \qw & \gate[2,disable auto height]{C^{\dagger}}\vqw{-1} & \qw & \gate[2,disable auto height]{C}\vqw{-1} & \qw & \gate[2,disable auto height]{C^{\dagger}}\vqw{-1} & \qw & \qw\\
\qw & \qw & \qw & \qw & \qw & \qw & \qw & \qw & \qw & \qw & \qw & \qw & \qw & \qw & \qw & \qw & \qw & \qw & \qw & \qw & \qw & \qw & \qw & \qw & \qw & \qw & \qw & \qw & \qw & \qw & \qw & \qw & \qw & \qw
\end{quantikz}
}
\caption{\label{dqc1_circ} Schematic of the $16$-point OTOC circuit described by $\Phi_{16}[C]$ that estimates normalised trace of circuit $C$.}
\end{figure}
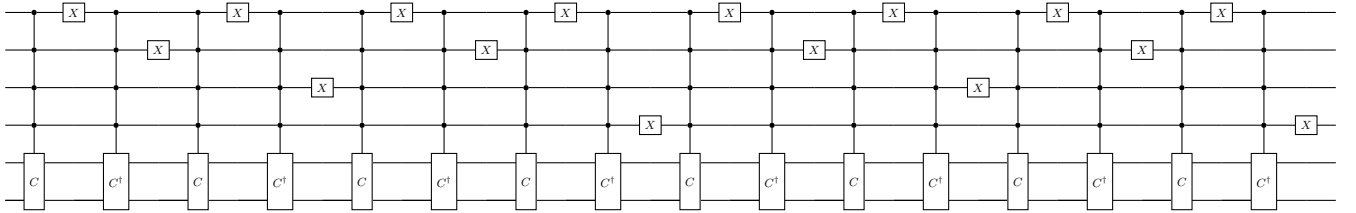

\end{example}

\end{document}